\documentclass[
aps,prd,
preprintnumbers,
amsmath,
amssymb,
nofootinbib,
preprintnumbers
]{revtex4}
\usepackage{here}
\usepackage{footnote}
\usepackage{comment,braket}
\usepackage{epsf}
\usepackage{amsmath}
\usepackage{graphics}
\usepackage{amsfonts}
\usepackage{amssymb}
\usepackage{latexsym}
\usepackage{color}
\usepackage{natbib}
\usepackage{graphicx}
\usepackage{hyperref}
\usepackage{array}
\usepackage{amsthm}

\newtheorem*{theorem-main}{The CGM Theorem}
\newtheorem*{theorem-new}{The Main Theorem}
\newtheorem*{theorem-a}{Theorem A}
\newtheorem*{theorem-b}{Theorem B}
\newtheorem*{definition}{Definition}
\newtheorem{proposition}{Proposition}

\newcommand{\cft}{\text{CFT}}
\newcommand{\ads}{\text{AdS}}
\newcommand{\bdy}{\text{bdy}}

\usepackage[svgnames]{xcolor}
\definecolor{phthaloblue}{rgb}{0.0, 0.06, 0.54}
\hypersetup{
    colorlinks=true,
    linkcolor=blue,
    citecolor=blue,
    filecolor=blue,
    urlcolor=blue,
    }

\begin{document}

\title{Is the Coleman de Luccia action minimum?: AdS/CFT approach}
%
\author{Naritaka Oshita$^{1,2,3}$}
\author{Yutaro Shoji$^4$}
\author{Masahide Yamaguchi$^{5,6}$}
\preprint{YITP-23-99, RIKEN-iTHEMS-Report-23}
\affiliation{$^1$Center for Gravitational Physics, Yukawa Institute for Theoretical Physics,
Kyoto University, Kitashirakawa Oiwakecho, Sakyo-ku, Kyoto 606-8502, Japan}
\affiliation{$^2$The Hakubi Center for Advanced Research, Kyoto University,
Yoshida Ushinomiyacho, Sakyo-ku, Kyoto 606-8501, Japan}
\affiliation{$^3$RIKEN iTHEMS, Wako, Saitama, 351-0198, Japan}

\affiliation{
 $^4$Racah Institute of Physics, Hebrew University of Jerusalem, Jerusalem 91904, Israel
}
\affiliation{
 $^5$Cosmology, Gravity and Astroparticle Physics Group,\\ Center for Theoretical Physics of the Universe, Institute for Basic Science (IBS), Daejeon, 34126, Korea
}
\affiliation{
  $^6$Department of Physics, Tokyo Institute of Technology, 2-12-1 Ookayama, Meguro-ku, Tokyo 152-8551, Japan
}

\begin{abstract}
We use the anti-de Sitter/conformal field theory (AdS/CFT) correspondence to find the least bounce action in an AdS false vacuum state, i.e., the most probable decay process of the metastable AdS vacuum within the Euclidean formalism by Callan and Coleman. It was shown that the $O(4)$ symmetric bounce solution leads to the action minimum in the absence of gravity, but it is non-trivial in the presence of gravity. The AdS/CFT duality is used to evade the difficulties particular to a metastable gravitational system, such as the problems of negative modes and unbounded action. To this end, we show that the Fubini bounce solution in CFT, corresponding to the Coleman de Luccia bounce in AdS, gives the least action among all finite bounce solutions in a conformal scalar field theory.
Thus, we prove that the Coleman de Luccia action is the least action when (i) the background is AdS, (ii) the AdS radii, $L_+$ and $L_-$, in the false and true vacua, respectively, satisfy $L_+ / L_- \simeq 1$, and (iii) a metastable potential gives a thin-wall bounce much larger than the AdS radii.
\end{abstract}

\maketitle

\section{Introduction}
The vacuum decay process can be important both in the early and later Universe. In the early universe, vacuum decay may lead to the graceful exit of the open inflation \cite{Gott:1982zf,Gott:1984ps,Sasaki:1993ha}. In the later Universe, the possible Higgs metastability \cite{Degrassi:2012ry,Buttazzo:2013uya}, predicted in the particle physics, would eventually lead to the nucleation of a negative-energy vacuum bubble and destroy the structure of the present Universe. Also, the string theory predicts the existence of many vacuum states with various values of cosmological constants, which is known as the string landscape \cite{Susskind:2003kw}. In the picture of landscape, a universe could have various cosmological constants by experiencing vacuum decay.

To quantify the decay rate $\Gamma$, we consider the Euclidean path integral under the semi-classical approximation and obtain $\Gamma = A e^{-B}$ from the bounce solution \cite{Coleman:1977py,Callan:1977pt}, where $A$ is a pre-factor and $B$ is the on-shell Eucldiean action of the bounce. The pre-factor $A$ can be estimated by the energy scale of a metastable system and the exponent $B$ governs the order of magnitude of the decay rate. Therefore, determining the factor $B$ is rather important to estimate the probability of a vacuum decay. Finding the most probable process among all possible processes is equivalent to finding the least Euclidean action among all possible bounce solutions in the Euclidean formalism. In the absence of gravity, Coleman, Glaser, and Martin (CGM) has proven \cite{Coleman:1977th} that the $O(4)$-symmetric vacuum bubble leads to the least action under some conditions. However, with gravity, there exist serious issues, e.g., the negative mode problem \cite{Lavrelashvili:1985vn,Tanaka:1992zw} and unboundedness problem \cite{Gibbons:1978ac}, and it is non-trivial if the maximally symmetric non-trivial solution, i.e., an $O(4)$ bounce, leads to the minimum action in the existence of gravity. For the vacuum decay processes that we are interested in, gravity can be strong and one cannot get rid of the degrees of freedom of gravity from the system. In this sense, we could say that the theory of vacuum decay has been facing the aforementioned serious issues.

\begin{figure}[t]
  \begin{center}
    \includegraphics[keepaspectratio=true,height=60mm]{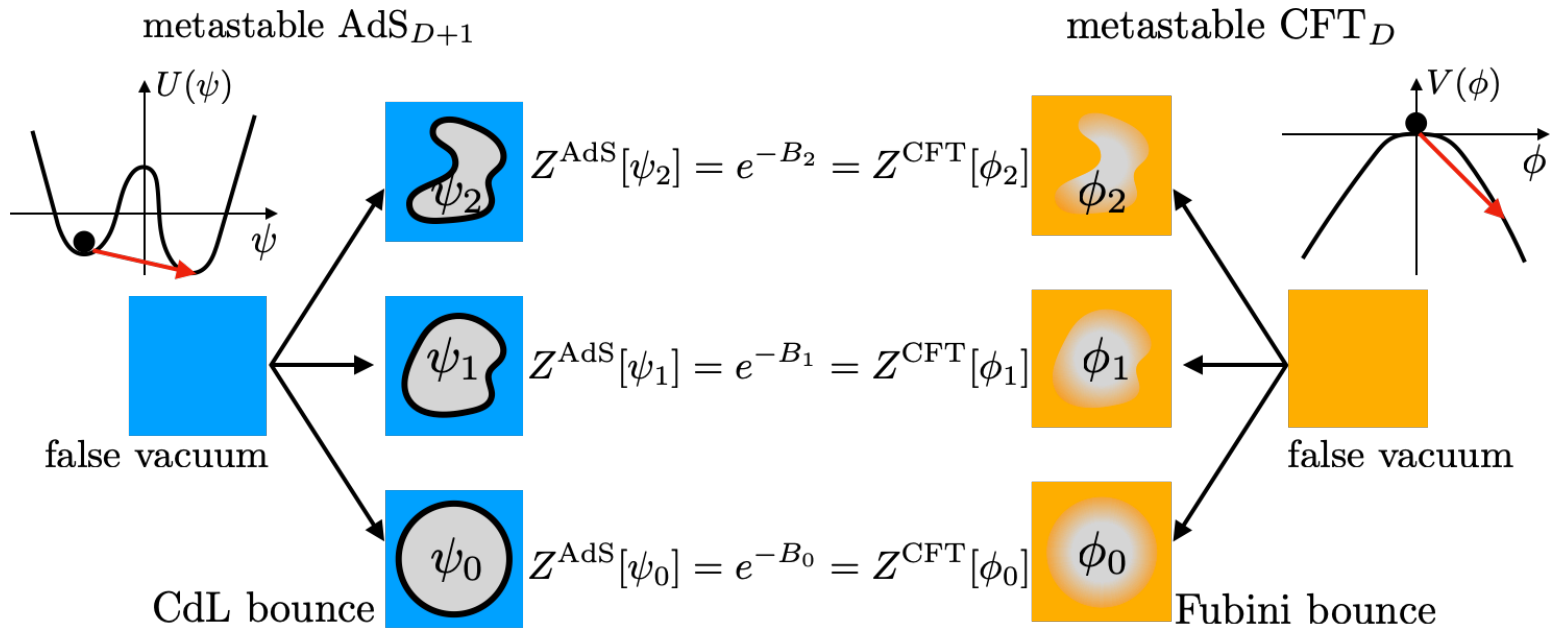}
  \end{center}
\caption{A schematic picture showing the role of the AdS/CFT correspondence in our strategy to find the least bounce action in the presence of gravity.
}
\label{pic_ads_cft}
\end{figure}

We consider how the anti-de Sitter/conformal field theory (AdS/CFT) correspondence \cite{Maldacena:1997re,Gubser:1998bc,Witten:1998qj} can shed light on the issues. We assume that the correspondence holds for a metastable AdS and CFT, that is, there exists a one-to-one correspondence between the partition functions of a bounce solution in AdS and CFT sides. We then find the least action in the CFT side where gravity is absent, which infers what is the least action in AdS side by virtue of the AdS/CFT correspondence (see Figure \ref{pic_ads_cft}). As mentioned, finding the least action among possible bounce solutions in the presence of gravity is challenging, but we can use the AdS/CFT correspondence to evade the complicated issues caused by gravity. We will then argue that the CdL bounce would correspond to the Fubini bounce under certain conditions. We then prove that the CdL bounce in the CFT side is always spherically symmetric and hence it is given by the Fubini bounce. Knowing that the spherically symmetric thin-wall bounce in the AdS side gives the same action as the Fubini bounce in the CFT side, we conclude that the spherically symmetric bounce gives the least action in the AdS side under certain conditions.

This paper is organized as follow. In Sec.~\ref{sec_adscft}, we set up the condition under which our strategy works. We consider a metastable scalar field theory in AdS$_{D+1}$ and review a way of how to determine the corresponding CFT$_{D}$ with the correct coupling constant based on Ref.~\cite{Barbon:2010gn}. In Sec.~\ref{sec_fubini}, we prove that the Fubini bounce solution is the least action among possible finite non-trivial solutions to a metastable conformal scalar field theory. We then provide our conclusions in Sec.~\ref{sec_discussion}. Throughout the paper, we use the natural units with $c= \hbar = 1$ and $G=1$.

\section{Correspondence between a metastable AdS${}_{D+1}$ and CFT${}_D$}
\label{sec_adscft}
In this section, we consider the correspondence between a metastable AdS${}_{D+1}$ and a metastable CFT${}_D$:
\begin{align}
S_{{\rm AdS}_{D+1}} &= \int d^{D+1} x \sqrt{-g} \left( \frac{1}{16 \pi} {\cal R} - \frac{1}{2} g^{\mu \nu}\partial_{\mu} \psi \partial_{\nu} \psi -U(\psi) \right),\\
S_{\cft_D} &= \int d^D y \sqrt{-g_{\bdy}} \left( -\frac{1}{2} g_{\bdy}^{ij} \partial_{i} \phi \partial_{j} \phi -\frac{1}{2} \xi_D {\cal R}_{\bdy} \phi^2 - \lambda \phi^{2D/(D-2)} \right),
\label{cft3}
\end{align}
where $\xi_D \equiv \frac{D-2}{4(D-1)}$ (see e.g. Ref.~\cite{Jacobson:2003vx}), $U(\psi)$ is a metastable potential, and the coupling constant $\lambda$ will be determined later but is negative in order for the CFT to be a metastable system. Following the AdS/CFT correspondence, we assume that there is a one-to-one correspondence between bounce solutions $\psi= \bar{\psi}$ in AdS$_{D+1}$ and $\phi= \bar{\phi}$ in CFT$_{D}$ such that
\begin{equation}
Z_{{\rm AdS}_{D+1}} (\bar{\psi}) = Z_{{\rm CFT}_{D}} (\bar{\phi}),
\end{equation}
where the left (right) hand side is the partition function of a bounce solution nucleated in the bulk (on the boundary). As the partition functions of the bounce and the initial false vacuum determine the transition amplitude in the Euclidean path integral formalism, we may expect that the corresponding bubbles in AdS and CFT sides would be nucleated with the same transition amplitude. Such a one-to-one correspondence means that the transition amplitude of the most probable decay process in the metastable AdS is equivalent to that in the metastable CFT side. Our goal in this paper is to confirm that, with this assumption, the CdL nucleation process is the most probable process at least in the AdS background.

Given the metastable potential in AdS, $U(\psi)$, how can we determine the coupling constant $\lambda$ in CFT? We can demonstrate the determination of $\lambda$ from the AdS side when the $U(\psi)$ satisfies the conditions shown below.
We here consider a metastable potential $U(\psi)$ for which the $O(D+1)$ bounce solution, i.e., the CdL solution, has a large wall with the exterior and the interior AdS radii, $L_+$ and $L_-$, respectively, and a potential barrier of the tension $\sigma$ such that
\begin{equation}
0< q/\sigma -1 \ll 1 \ \text{and} \ L_+/L_- \simeq 1,
\label{large_bubble_condition}
\end{equation}
where
\begin{equation}
q \equiv \frac{(L_+^2/L_-^2-1)-L_+^2 \Sigma^2}{16 \pi L_+/(D-1)} \ \text{and} \ \Sigma \equiv 8 \pi \sigma/(D-1).
\label{large_bubble_cond}
\end{equation}
Here the tension of the wall is given by $\sigma \sim \sqrt{V_{\rm top}} \Delta \phi$ where $\Delta \phi$ is the separation of the true and false vacuum states in the field space and $V_{\rm top}$ is the height of the potential barrier $V_{\rm top}$. The two quantity $q$ and $\sigma$ are associated with the bulk and surface energy, respectively, and the balance between them determines the size of the CdL bubble.
In the condition (\ref{large_bubble_cond}), one of the possible bounce solutions, the CdL solution, has the bubble radius of $R_0$ which is much larger than the false AdS radius as
\begin{equation}
R_0 \equiv \frac{\sigma}{\sqrt{q^2-\sigma^2}} L_+ \gg L_+,
\end{equation}
where the explicit form of $R_0$ is derived below.
For the CdL solution, all degrees of freedom in the bulk, i.e., the thin wall or a probe brane, lives in the vicinity of the AdS boundary, and its dynamics can be translated into that of CFT \cite{Barbon:2010gn}. Then one can read the unknown coupling constant $\lambda$ from the bulk side by sending the probe brane to the vicinity of the AdS boundary at $r \gg L_+$\footnote{The coordinate $r$ is the radial coordinate in the static AdS patch.} and obtaining the effective action of the probe brane in the canonical form \cite{Barbon:2010gn}. In the following, we review the procedure of Ref.~\cite{Barbon:2010gn}.

The dynamics of a thin-wall spherical bubble can be described by the Israel junction conditions \cite{Israel:1966rt}. It means that the effective degrees of freedom of the bulk reduces to a scalar quantity, i.e., the radius of the bubble. As we consider a spherical probe brane, the first Israel junction condition is trivially satisfied and the second Israel junction condition reduces to
\begin{align}
&\frac{\sqrt{f_+ + (dR/d\tau)^2}}{R} - \frac{\sqrt{f_- + (dR/d\tau)^2}}{R} = -\Sigma,
\label{IJ_2nd}\\
&f_{\pm} \equiv 1+R^2/L_{\pm}^2,
\end{align}
where $r = R(\tau)$ denotes the radius of the brane and $\tau$ is its proper time. The junction condition (\ref{IJ_2nd}) reduces to
\begin{align}
&\left(\frac{dR}{d\tau} \right)^2 + 1- \frac{q^2- \sigma^2}{\sigma^2 L_+^2} R^2 = 0.
\label{canonical_form}
\end{align}
Note that $q > 0$ should hold for the positivity of the exterior and interior extrinsic curvatures. From (\ref{canonical_form}), we find the radius at the moment of the bubble nucleation is $R = R_0$, at which the potential term in (\ref{canonical_form}) becomes zero, and $R_0 \to \infty$ for $\sigma \to q$. Using the asymptotic time, $t$, \eqref{canonical_form} can be rewritten as
\begin{equation}
\left(\frac{dR}{dt} \right)^2 + f_+^2 \left( \frac{f_+ \sigma^2 L_+^2}{q^2 R^2} -1 \right) = 0.
\label{canonical_form_asmptotic}
\end{equation}
The action leading to the integrated equation of motion (\ref{canonical_form_asmptotic}) is given by
\begin{equation}
S_{\ads} = \int dt L = -\sigma \Omega_{D-1} \int dt R^{D-1} \sqrt{f_+ - \frac{\dot{R}^2}{f_+}} + \frac{q}{L_+} \Omega_{D-1} \int dt R^D,
\label{ads_system}
\end{equation}
where $\Omega_{D-1}$ denotes the area of the $(D-1)$-dimensional unit sphere and a dot denotes the derivative with respect to $t$. One can show that the action (\ref{ads_system}) indeed derives the integrated equation of motion (\ref{canonical_form_asmptotic}) by computing the Hamiltonian (total energy) of the bubble $E$ as
\begin{align}
E &= \dot{R} \frac{\partial L}{\partial \dot{R}} - L 
= \frac{\sigma \Omega_{D-1} R^{D-1} f_+}{\sqrt{f_+ - \dot{R}^2/f_+}} - \frac{q}{L_+} \Omega_{D-1} R^D,
\label{bubble_energy}
\end{align}
and setting $E=0$ as the total energy of the nucleated bubble is zero, one finds that (\ref{bubble_energy}) reduces to (\ref{canonical_form_asmptotic}). In the following, we obtain the translation of $R(t) \to \phi(t)$, by which the action $(\ref{ads_system})$ reduces to the canonical form of 
\begin{equation}
L = L_+^{D-1} \Omega_{D-1} \left(\frac12 \dot{\phi}^2 - V_{\ads} (\phi) + {\cal O}(\dot{\phi}^4) \right),
\label{canonical_form_cft}
\end{equation}
in the non-relativistic situation $\dot{R} \ll 1$. This is the case when the bubble is nucleated with a small velocity\footnote{Indeed, the bubble is nucleated with $\dot R=0$ in the CdL formalism.}. In this procedure, we can read the $\lambda$ in CFT from the bulk side. Expanding $L (\dot{R}, R)$ in (\ref{ads_system}) with respect to $\dot{R}$ and comparing it with (\ref{canonical_form_cft}), one can read
\begin{equation}
\phi = L_+^{\frac{4-D}{2}} \frac{2\sqrt{\sigma}}{D-2} R^{(D-2)/2} (1+ {\cal O}(L_+^2/R^2)),
\end{equation}
and substituting this relation and $\dot{R} = 0$ in (\ref{bubble_energy}), one finds
\begin{align}
V_{\ads} (\phi) = \sigma\, (R(\phi)/L_+)^{D-1} \sqrt{f_+(R(\phi))} -q\, (R(\phi)/L_+)^D
\simeq \frac{(D-2)^2}{8} (\phi/L_+)^2 + \lambda_{\ads} \phi^{2D/(D-2)},
\label{bulk_potential}
\end{align}
for $R \gg L_+$, where
\begin{equation}
\lambda_{\ads} \equiv -\left( \frac{D-2}{2} \right)^{\frac{2D}{D-2}} \frac{1}{\sigma^{\frac{2}{D-2}}} \frac{q-\sigma}{\sigma} \frac{1}{L_+^{2D/(D-2)}}.
\end{equation}
The potential term (\ref{bulk_potential}) reduces to
\begin{equation}
V_{\ads} (\phi) \simeq \frac{1}{2} \xi_{\rm D} {\cal R}_{\bdy} \phi^2 + \lambda_{\ads} \phi^{2D/(D-2)},\label{ads_potential}
\end{equation}
as the Ricci scalar is ${\cal R}_{\bdy} = (D-1) (D-2)/L_+^2$ on the AdS boundary whose topology is ${\bf R} \times {\bf S}^{D-1}$. Identifying $\lambda$ in (\ref{cft3}) with $\lambda_{\rm AdS}$, we obtain the CFT action corresponding to the metastable AdS satisfying the condition of (\ref{large_bubble_condition}).
Let us consider the correspondence in the Wick rotated space ($t \to - i t_{\rm E}$) and perform the conformal transformation leading to ${\bf R} \times {\bf S}^{D-1} \to {\bf R}^D$. The latter procedure is possible as
\begin{equation}
dt_{\rm E}^2 + d\Omega_{D-1} = \frac{1}{u^2} (du^2 + u^2 d\Omega_{D-1}^2),
\end{equation}
where $u \equiv \exp(t_{\rm E})$ and the factor $1/u^2$ is the conformal factor of the tranformation. 
Performing the conformal transformation, ${\cal R}_{\rm bdy}$ vanishes and \eqref{ads_potential} becomes
\begin{equation}
V_{\ads} (\phi) \simeq V_{\rm CFT} (\phi) \equiv \lambda_{\ads} \phi^{2D/(D-2)}.\label{cft_potential}
\end{equation}
Then, the $D$-dimensional Fubini bounce \cite{Fubini:1976jm} becomes a solution to the equations of motion for (\ref{cft3}) with ${\cal R}_{\rm bdy} = 0$.
Here, the Fubini bounce is given by
\begin{equation}
    \phi = \left(\frac{2}{|\lambda|}\right)^{1/2} \left(\frac{\Delta b}{x^2+b^2}\right)^{\Delta},\label{fubini_bounce}
\end{equation}
where $\Delta = (D-2)/2$ is the mass dimension of a scalar field $\phi$ and $b$ is an arbitrary constant determining the size of the bounce.
As the conformal transformation does not affect the action, the original CFT of (\ref{cft3}) also admits the on-shell Fubini action.
Remarkably, the Fubini action with $\lambda = \lambda_{\rm AdS}$
\begin{equation}
S_{\rm Fubini} = \frac{c}{\lambda_{\rm AdS}^{\Delta}}, \ 
c \equiv \frac{\Omega_{D-1}}{2^{\Delta}} \Delta^{D} B\left( \frac{3}{2}, \frac{D-2}{2} \right),
\end{equation}
is equivalent to the Coleman de Luccia action in the limit of $L_+/L_- \to 1$ as it has the form of
\begin{equation}
S_{\rm AdS} = \frac{L_+}{L_-} S_{\rm Fubini},
\label{relation_Scdl_Sfubini}
\end{equation}
for $q / \sigma \to 1$. The exterior and interior AdS radii, $L_+$ and $L_-$, respectively, should satisfies $L_+/L_--1 \sim 1/N \ll 1$ for the AdS/CFT correspondence to be valid, where $N$ is a large integer. In the context of AdS/CFT, the $N$ is the number of branes, and for $N \gg 1$, the spacetime near the branes is approximated with AdS spacetime. Also, a nucleated bubble can be regarded as bundled $n$ branes with $n \ll N$ \cite{Barbon:2010gn}.

In the following section, we show that the Fubini bounce leads to the least bounce action among all finite bounce solutions by extending the theorem on the minimum action proven by Coleman, Glaser, and Martin \cite{Coleman:1977th} to cover a scalar CFT (see Sec. \ref{sec:CGM_extension}). Based on the relation of (\ref{relation_Scdl_Sfubini}), we argue that the CdL bounce action gives the most probable transition amplitude among all possible processes, at least, in our setup.

\section{Most probable decay process in the metastable CFT}
\label{sec_fubini}
\label{sec:CGM_extension}
In this section, we will prove that the Fubini bounce gives the least Euclidean action of the metastable CFT. To this end, we extend the theorem proven by Coleman, Glaser, and Martin \cite{Coleman:1977th} (hereinafter, we refer it as the CGM theorem). In the former part of this section, we will briefly review the CGM theorem, and in the latter part, we will extend the CGM theorem so that it applies to the Fubini bounce.

\subsection{CGM theorem}
CGM has shown that there exists at least one non-trivial solution to the differential equation
\begin{equation}
\nabla^2\phi-V'(\phi) = 0,
\label{Euclidean_nongravity_eom}
\end{equation}
and the solution leading to the lowest action is spherically symmetric and monotone if $V(\phi)$ is {\it admissible}. Here, $\nabla^2$ is the Laplacian in the $D$ dimensional Euclidean space $\{x_1, x_2, ..., x_D\}$, and $V(\phi)$ is said to be admissible if i) $V$ is continuously differentiable for all $\phi$, ii) $V(0) = V'(0) = 0$, iii) $V$ is somewhere negative, and iv) there exist positive numbers $a$, $b$, $\alpha$, and $\beta$ such that
\begin{equation}
\alpha < \beta < 2D/(D-2),
\label{albetD}
\end{equation}
with
\begin{equation}
V - a|\phi|^{\alpha} + b |\phi|^{\beta} \geq 0.
\label{potential_cond}
\end{equation}
Notice that this is not the case for the potential of \eqref{cft_potential} as we need $\beta = 2D/(D-2)$ and $a=0$ to satisfy the inequality.

The main theorem proven by CGM is:
\begin{theorem-main}
In $D$-dimensional Euclidean space with $D>2$, for any admissible $V$, the equation of motion (\ref{Euclidean_nongravity_eom}) has at least one monotone spherical solution vanishing at infinity, other than the trivial solution of $\phi = 0$. Furthermore, this solution has Euclidean action,
\begin{equation}
S = \int d^D x \left[ \frac{1}{2} (\partial \phi)^2 + V(\phi) \right], \label{euclidean_action}
\end{equation}
less than or equal to that of any other solution vanishing at infinity. If the other solution is not both spherical and monotone, the action is strictly less than that of the other solution.
\end{theorem-main}

Before proving the theorem, CGM defined the reduced problem as follows.
\begin{definition}
``The reduced problem" is the problem of finding a function vanishing at infinity which minimizes $T$ for some fixed negative $W$, where
\begin{equation}
T[\phi] \equiv \int d^Dx \frac{1}{2} (\partial \phi)^2, \ W[\phi] \equiv \int d^Dx V(\phi).
\label{def_T_W}
\end{equation}
\end{definition}
It is equivalently stated as the problem to minimize the scale-invariant ratio,
\begin{equation}
    X[\phi]=-\frac{(T[\phi])^{D/(D-2)}}{W[\phi]},
\end{equation}
with negative $W$.

The CGM theorem is proven by showing that the following theorems hold.
\begin{theorem-a}
If a solution of the reduced problem exists, then, for an appropriate value of $W$, it is a solution of (\ref{Euclidean_nongravity_eom}) that has an action less than or equal to that of any non-trivial solution of (\ref{Euclidean_nongravity_eom}).
\end{theorem-a}
\begin{theorem-b}
There exists at least one solution to the reduced problem. All solutions to the reduced problem are spherically symmetric and monotone.
\end{theorem-b}
The proof of Theorem B is composed of a sequence of statements with short proofs. CGM start from an infinite minimizing sequence, $\{\phi_n\}$, $n\in{\mathbb Z}_+$, such that
\begin{equation}
    \lim_{n\to\infty}T[\phi_n]=\inf_\phi T[\phi],
\end{equation}
with a fixed negative $W$.
The sequence is chosen so that $\phi_n$ is differentiable and has compact support, and $T[\phi_n]$ is finite. Notice that such a choice of the sequence is always possible. 
Then, CGM has proven the following statements.
\begin{enumerate}
    \item[(a).] [CGM Statement 4] There exists a sequence of spherical and monotone functions, $\{\phi_n^{\rm sph}\}$, such that 
    \begin{equation}
        X[\phi_n^{\rm sph}]\leq X[\phi_n],
    \end{equation}
    for all $n\in I$. Here, $I$ is an infinite subset of ${\mathbb Z}_+$.
    \item[(b).] [CGM pp.220-221] There is no non-spherical or non-monotone function that has the same $R$ as the spherical monotone rearrangement of the original function.
    \item[(c).] [CGM Statement 6] There exist an infinite subsequence, $\{\Phi_n\}$, of $\{\phi^{\rm sph}_n\}$ and a bounded continuous function, $\Phi(r)$, such that
    \begin{equation}
        \lim_{n\to\infty}\Phi_{n}(r)=\Phi(r),
    \end{equation}
    pointwise for all $r\in(0,\infty)$ and uniformly on any finite closed interval in $(0,\infty)$.
    
    \item[(d).] [CGM Statement 6]  $\Phi$ satisfies
    \begin{equation}
        \lim_{r\to\infty}\Phi(r)=0.
    \end{equation}
    \item[(e).] [CGM Statement 8] $\Phi$ satisfies
    \begin{equation}
        W[\Phi]<0.
    \end{equation}
    \item[(f).] [CGM Statement 10] $\Phi$ satisfies
    \begin{equation}
        X[\Phi]=\lim_{n\to\infty}X[\Phi_n].
    \end{equation}
\end{enumerate}
Here, (a) and (b) show that the solution to the reduced problem is always spherically symmetric and monotone, and (c)-(f) show the sequence converges to the actual minimum of $X$ satisfying $\lim_{r\to\infty}\Phi(r)=0$ and $W[\Phi]<0$. Hence, these statements prove Theorem B.

The other statements of CGM are used to prove Statements 4, 6, 8 and 10 shown above. The dependencies of the statements are summarized in Appendix \ref{apx_statements}.
\subsection{Extension of the CGM Theorem}
We consider non-trivial solutions to the differential equation (\ref{Euclidean_nongravity_eom}) with the potential\footnote{Addition of dimensionful terms, such as a mass term in $D=4$, potentially results in non-existence of the bounce as discussed in \cite{Frishman:1978xs,Affleck:1980mp,Nielsen:1999vq,DiLuzio:2015iua}.} of
\begin{equation}
V = \lambda \phi^{\gamma},
\label{potential_cond_2}
\end{equation}
where $\gamma = 2 D/(D-2)$ and $\lambda$ is a negative constant.
The theorem we here prove is described below.
\begin{theorem-new}
In $D$-dimensional Euclidean space with $D>2$,
the equation of motion (\ref{Euclidean_nongravity_eom}) with the potential of (\ref{potential_cond_2}) has at least one monotone spherical solution vanishing at infinity, other than the trivial solution of $\phi = 0$. Furthermore, the solution has the Euclidean action (\ref{euclidean_action}),
which is less than or equal to that of any other solution vanishing at infinity. If the other solution is not both spherical and monotone, the action is strictly less than that of the other solution.
\end{theorem-new}

To prove the Main Theorem, we show that Theorem A and B holds in our setup. Theorem A has been proven without the condition of \eqref{potential_cond} and thus our main focus is on Theorem B. As we have mentioned, Theorem B has been proven by showing (b) and Statements 4, 6, 8 and 10. As summarized in Appendix \ref{apx_statements}, (b) and Statements 4 and 6 hold independently of \eqref{potential_cond}, and Statement 10 follows from Statement 8.
However, the proof of Statement 8 by CGM depends on \eqref{potential_cond} and does not apply to our case.
This can be understood in the following way. Since $V(\phi)<0$ for any $\phi\neq0$, there is a possibility that a sequence of $\Phi_n$ having a fixed negative $W$ converges to $\Phi$ that is zero almost everywhere. If such a sequence exists, we obtain $W[\Phi]=0$ although $W[\Phi_n]<0$ for all $n$, which contradicts Statement 8. In fact, we can construct such a sequence utilizing the scale invariance of the theory. (One can see that any value of $b$ in \eqref{fubini_bounce} gives the same bounce action, which is the consequence of the scale invariance.) For any sequence that converges to the Fubini bounce, we can execute the scale transformation at each step of $n$ so that the new sequence converges to the Fubini bounce with $b\to 0$ or $b\to\infty$, which is zero almost everywhere.

Let us move on to the proof of the Main Theorem. Since the proof of Statement 4 by CGM applies to our setup, there exists a minimizing sequence, $\{\phi_n\}$, such that $\phi_n$ is spherically symmetric and monotone for all $n$.
Hereafter, we write the functions in terms of the radius from the center, $r=e^y$.

We prove the following propositions.
\begin{proposition}
    There exists a minimizing sequence of spherically symmetric monotone functions, $\{\Phi_n\}$, such that (i) $f_n(y)=\Phi_n(e^y) e^{\frac{D-2}{2}y}$ is symmetric under $y\to-y$ and monotone for $y>0$ for all $n$, and (ii) there exists a bounded continuous function, $f(y)$, such that
    \begin{equation}
        \lim_{n\to\infty}f_n(y)=f(y),
    \end{equation}
    pointwise for all $y$ and uniformly on any finite interval.
\end{proposition}
\begin{proposition}[Statement 8']
    For the minimizing sequence of the preceeding proposition,
    \begin{equation}
        W[\Phi]=W[\Phi_n].
    \end{equation}
\end{proposition}
Since the scale transformation corresponds to the translation in $y$ space, Proposition 1 excludes the sequences that converge to the Fubini bounce with $b\to0$ or $b\to\infty$. Then, Proposition 2 replaces the Statement 8 of CGM.
Once Proposition 2 is proven, Statement 10 of CGM immediately follows from Proposition 2 and Statement 9 of CGM, which completes the proof of Theorem B and the Main Theorem.

\begin{figure}[t]
  \begin{center}
    \includegraphics[keepaspectratio=true,height=100mm]{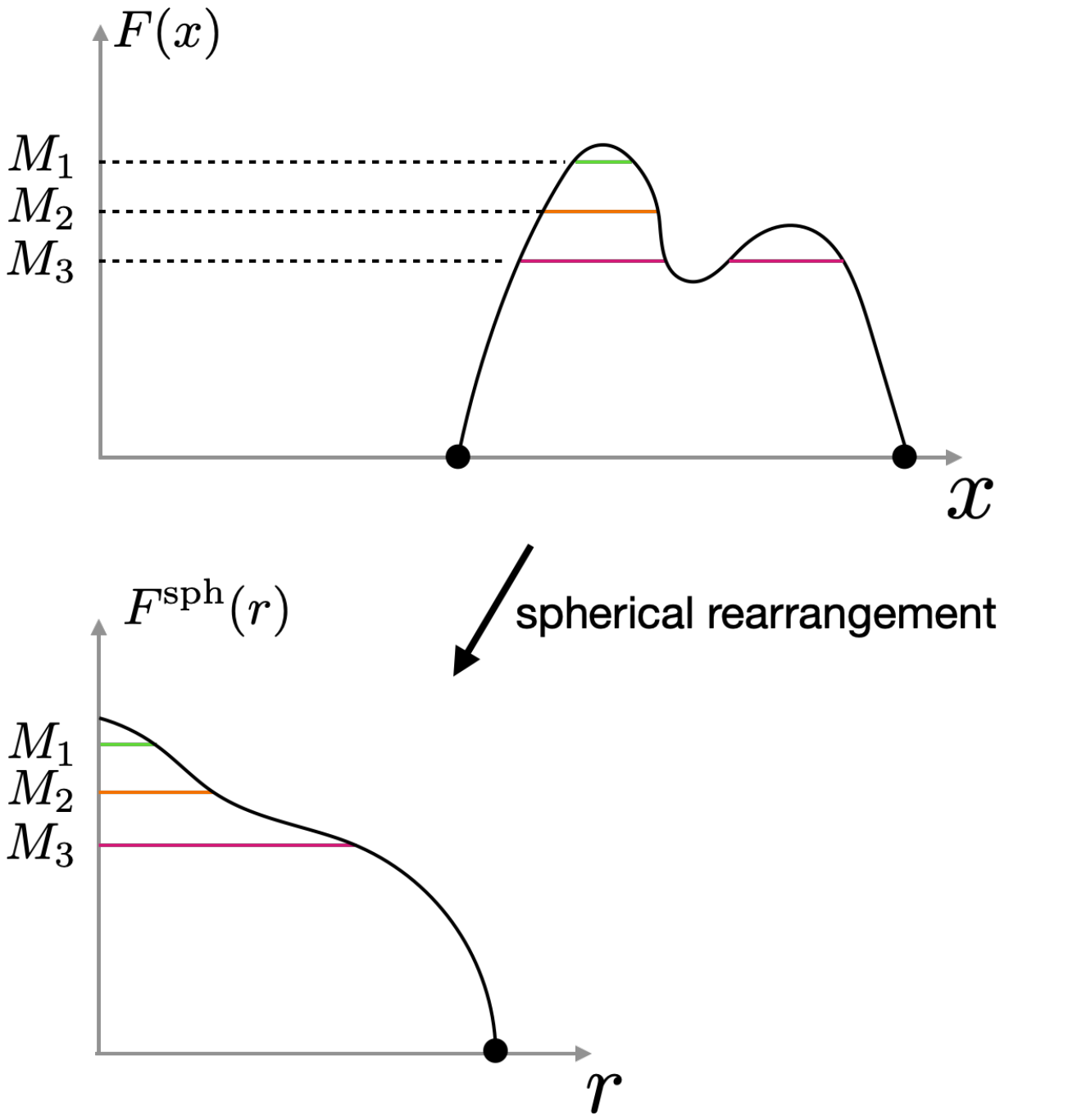}
  \end{center}
\caption{A schematic picture of the spherical rearrangement $F^{\rm sph} (x)$. The area of the level set of $F = M_i$ ($i=1,2,3$) is equal to that of $F^{\rm sph} = M_i$.}
\label{pic_spherical_rearrangement}
\end{figure}

\begin{definition}[Spherical rearrangement]
Let $F(x)$ be a non-negative measurable function on ${\mathbb R}^d$ $(d\geq1)$ that vanishes at infinity.
A spherically symmetric monotone function, $F^{\rm sph}(r)$, is obtained by symmetrizing $F(x)$ around $r\equiv \sqrt{x_1^2 +x_2^2 + ... + x_d^2}=0$ keeping
\begin{equation}
{\cal A} \left\{ x| F^{\rm sph}(r) \geq M \right\}
= {\cal A} \left\{ x| F(x) \geq M \right\},
\end{equation}
for any positive value $M$. Here, ${\cal A}$ is the Lebesgue measure.
Then, $F^{\rm sph}(r)$ is said to be a spherical rearrangement of $F(x)$.  (See Figure \ref{pic_spherical_rearrangement}.)
\end{definition}
The spherical rearrangement has the following properties.
\begin{align}
    ||F^{\rm sph}||_{L^p}&=||F||_{L^p},\label{spherical_rearrangement_prop1}\\
    ||\nabla F^{\rm sph}||_{L^p}&\leq||\nabla F||_{L^p},\label{spherical_rearrangement_prop2}
\end{align}
where $||*||_{L^p}$ is the $L^p$-norm with $1\leq p<\infty$.

\begin{proof}[proof of Prop 1]
    With $\tilde f_n=\phi_n(e^y) e^{\frac{D-2}{2}y}$, $W$ and $T$ can be rewritten as
    \begin{align}
        W[\tilde f_n]&=\Omega_{D-1}\int_{-\infty}^\infty dy\lambda \tilde f_n^\gamma,\\
        T[\tilde f_n]&=\Omega_{D-1}\int_{-\infty}^\infty dy\left[\frac12\tilde f'^2_n+\frac{(D-2)^2}{8}\tilde f_n^2\right].
    \end{align}
    
    Let $\tilde f_n^{\rm sph}$ be the spherical rearrangement of $\tilde f_n$ in $y$ space.
    Then, from \eqref{spherical_rearrangement_prop1} and \eqref{spherical_rearrangement_prop2},
    \begin{align}
        W[\tilde f_n^{\rm sph}]=W[\tilde f_n],\\
        T[\tilde f_n^{\rm sph}]\leq T[\tilde f_n].
    \end{align}
    Thus, there exists a subsequence of $\{\tilde f_n^{\rm sph}\}$ that is a minimizing sequence satisfying property (i).
    Then, the sequence with property (ii) is obtained by applying Statement 6 of CGM to this sequence.
\end{proof}
\begin{proof}[proof of Prop 2]
    Statement 5 (B) of CGM holds in our setup: there exists $M>0$ such that
    \begin{equation}
        \int_0^\infty dyf_n^{2}(y)\leq M.\label{statement_5_B}
    \end{equation}
    
    Let us take an arbitrary $y_1>0$. Since $|f_n(y)|$ is a non-negative monotonically decreasing function for $y>0$, we have
    \begin{equation}
        \frac{|W|}{2\Omega_{D-1}|\lambda|}\geq \int_0^{y_1}dy |f_n(y)|^\gamma\geq |f_n(y_1)|^\gamma y_1,
    \end{equation}
    Since $W[f_n]$ is independent of $n$, $|f_n(y_1)|$ is uniformly bounded as
    \begin{equation}
        |f_n(y_1)|\leq \left(\frac{|W|}{2\Omega_{D-1}|\lambda|y_1}\right)^{1/\gamma}.\label{f_n(y_1)}
    \end{equation}
    From \eqref{statement_5_B} and \eqref{f_n(y_1)}, it follows that
    \begin{align}
        \int_{y_1}^\infty dy |f_n(y)|^\gamma&\leq |f_n(y_1)|^{\gamma-2}\int_{y_1}^\infty dyf_n^{2}(y)\nonumber\\
        &\leq \left(\frac{|W|}{2\Omega_{D-1}|\lambda|y_1}\right)^{2/D}M.\label{integral_above}
    \end{align}
    Thus, this integral converges to zero uniformly as $y_1\to\infty$.

    Since $\lim_{n\to\infty}f_n$ converges to $f$ uniformly on $y\in[0,y_1]$, we have
    \begin{equation}
        \lim_{n\to\infty}\int_0^{y_1}dyf_n^\gamma (y)=\int_0^{y_1}dyf^\gamma (y).\label{integral_below}
    \end{equation}
    
    From \eqref{integral_above} and \eqref{integral_below}, it follows that
    \begin{equation}
        W[\Phi]=W[\Phi_n].
    \end{equation}
\end{proof}

We now know that the Fubini bounce solution, which is the spherical monotone bounce solution of the conformal scalar field theory, gives the least action among all finite non-trivial solutions.

\section{Discussion and conclusion}
\label{sec_discussion}
We here showed that the Fubini bounce is a solution leading to a minimum of the Euclidean action among all finite solutions to a conformal scalar field theory. Our proof is based on the proof in Ref.~\cite{Coleman:1977th}, which did not cover the conformal scalar field theory.

The action of a metastable conformal scalar field theory, corresponding to a metastable AdS, is uniquely determined except for the value of the coupling constant $\lambda$.
Assuming the one-to-one correspondence between the partition function of a metastable AdS bulk and that of the corresponding metastable CFT, we consider a case where the coupling constant of the CFT can be determined with $\lambda = \lambda_{\rm AdS}$ from the dynamics of a large bubble in the AdS.
Such a special situation is realized when a metastable effective potential of the bulk theory admits the nucleation of the CdL bubble in the vicinity of the AdS boundary.

It was shown \cite{Barbon:2010gn} that the Fubini solution with the coupling constant $\lambda_{\rm AdS}$ has the Euclidean action equivalent to the CdL action when $L_+/L_- \simeq 1$. We then conclude that the CdL action is the action minima among all possible finite solutions under the conditions we have discussed, provided that there exists the one-to-one correspondence between the metastable AdS and CFT.

The situation we investigated is restrictive and many open issues are remained: i) How does our procedure work for a case where a small CdL bubble is nucleated? ii) Is the CdL action minimum when the difference between the exterior and interior cosmological constants is not negligible? (i.e., $L_+/L_--1 \gtrsim 1$) iii) Is it possible to generalize our procedure to the case of Minkowski or de Sitter background? Beyond the Euclidean path integral technique, it is non-trivial that finding a solution leading to the least Euclidean action corresponds to finding the most probable process in vacuum decay. In Ref.~\cite{Oshita:2021aux,Shoji:2022rke}, the authors indeed considered another scheme, the polychronic tunneling (or mixed tunneling), where the Euclidean and Lorentzian evolution coexist during a phase transition. The authors then found that the polychronic tunneling is more probable than the CdL process at least when the false vacuum is almost Minkowskian.

The Euclidean path integral in the existence of gravity has many open issues that are challenging to address, e.g., unbounded-ness of the action, negative mode problems, and the choice of vacuum state. As such, it is totally non-trivial if the CdL action is one of the action minima among all possible bounce solutions. Our strategy to attack this problem is based on the one-to-one correspondence between the theory including gravity (AdS) and one without gravity (CFT). As mentioned, this is the first step towards solving the problem. As the vacuum decay process with strong gravity plays an important role in cosmology, the aforementioned open issues are very important. We would also expect that the AdS/CFT correspondence and other possible dualities (e.g., dS/CFT correspondence \cite{Strominger:2001pn}) could be helpful to understand vacuum decay.

\appendix
\section{The Structure of CGM Statements}
\label{apx_statements}
The CGM statements 4, 6, 8 and 10 are proven by other statements of CGM, and (b) [CGM pp.220-221] is an independent statement proven without \eqref{potential_cond}. Here, we summarize the dependencies of the CGM statements.
\begin{itemize}
    \item Statement 4
    \begin{itemize}
        \item Statement 3
    \end{itemize}
    \item Statement 6
    \begin{itemize}
        \item Statement 5 (C)
        \item Statement 5 (D)
    \end{itemize}
    \item Statement 8 $\clubsuit$
    \begin{itemize}
        \item Statement 7 $\clubsuit$
        \begin{itemize}
            \item Statement 5 (D)
            \item Statement 5 (F) $\clubsuit$
            \item Statement 6
        \end{itemize}
    \end{itemize}
    \item Statement 10 $\clubsuit$
    \begin{itemize}
        \item Statement 8 $\clubsuit$
        \item Statement 9
    \end{itemize}
\end{itemize}
The parent bullet depends on the child bullets, and $\clubsuit$ indicates that the statement requires \eqref{potential_cond}. For the details of each statement, see \cite{Coleman:1977th}.

Statement 5 (C), (D) and (F) further depend on other statements as follows.
\begin{itemize}
    \item Statement 5 (C)
    \begin{itemize}
        \item Statement 5 (A)
    \end{itemize}
        \item Statement 5 (D)
    \begin{itemize}
        \item Statement 5 (A)
        \item Statement 5 (B)
    \end{itemize}
     \item Statement 5 (F) $\clubsuit$
    \begin{itemize}
        \item Statement 2 $\clubsuit$
        \begin{itemize}
            \item Statement 1 $\clubsuit$
        \end{itemize}
        \item Statement 5 (E)
    \begin{itemize}
        \item Statement 5 (B)
        \item Statement 5 (D)
    \end{itemize}
    \end{itemize}
\end{itemize}

\begin{acknowledgements}
N.O. is supported by Grant-in-Aid for Scientific Research (KAKENHI) project for FY2023 (23K13111).
Y.S. is supported by the US-Israeli Binational Science Foundation (grant No. 2020220) and the Israel Science Foundation (grant No. 1818/22). 
M.Y. is supported by IBS under the project code, IBS-R018-D3, and by JSPS Grant-in-Aid for Scientific Research Number JP21H01080.
\end{acknowledgements}

\end{document}